\documentclass[aps,prd,groupedaddress,showpacs,showkeys]{revtex4}
\usepackage{epsfig}
\newcommand{\eq}{\begin{eqnarray}}
\newcommand{\en}{\end{eqnarray}}
\def\la{\langle}
\def\ra{\rangle}

\def\rp{$R_p\hspace{-1em}/\ \ $}
\def\rpp{$R_p\hspace{-0.8em}/\ \ $}

\begin{document}

\title{Implications of R-parity violating 
supersymmetry for atomic and hadronic EDMs} 
\author{ 
Amand Faessler$^1$,   
Thomas Gutsche$^1$,   
Sergey Kovalenko$^2$,      
Valery E. Lyubovitskij$^1$ 
\footnote{On leave of absence 
from Department of Physics, Tomsk State University, 
634050 Tomsk, Russia}
\vspace*{1.2\baselineskip}}

\affiliation{$^1$ Institut f\"ur Theoretische Physik,
Universit\"at T\"ubingen,
\\ Auf der Morgenstelle 14, D-72076 T\"ubingen, Germany
\vspace*{1.2\baselineskip} \\
\hspace*{-1cm}$^2$ Departamento de F\'\i sica, Universidad
T\'ecnica Federico Santa Mar\'\i a, \\
Casilla 110-V, Valpara\'\i so, Chile
\vspace*{0.3\baselineskip}\\}

\date{\today}

\begin{abstract}
We calculate the electric dipole moments (EDM) of the neutral 
$^{199}$Hg atom, deuteron, nucleons and neutral 
hyperons $\Lambda$, $\Sigma^0$ and $\Xi^0$ in the framework of 
a generic SUSY model without R-parity conservation (\rp SUSY) on the 
basis of the SU(3) version of chiral perturbation theory (ChPT). 
We consider CP-violation in the hadronic 
sector induced by the chromoelectric quark dipole moments 
and CP-violating 4-quark effective interactions. 
From the null experimental results on the neutron and  
$^{199}$Hg atom EDMs we derive limits on the imaginary parts of 
certain products Im$(\lambda^{\prime}\lambda^{\prime\ast})$ of the 
trilinear \rp-couplings and demonstrate that they are more stringent
than those existing in the literature. Using these limits we 
give predictions for the EDMs of neutral hyperons. 

We also estimate the prospects of future storage ring experiments 
on the deuteron EDM and show that the expected improvement of the above 
limits in these experiments may reach several orders of magnitude. 
\end{abstract}



\maketitle

\newpage 

\section{Introduction}

Electric dipole moments (EDMs) of neutral atoms, hadrons and leptons are  
flavor blind CP-odd observables which are recognized to be sensitive  
probes of physics beyond the Standard Model (SM). As is known, the SM   
predictions for these observables are at least 6-7 orders of magnitude 
below the current experimental limits. 
Thus an observation of EDMs at larger values would witness the presence 
of a non-SM source of CP-violation (CPV).   
Physics beyond the SM brings in new complex parameters and, therefore,  
new sources of CPV which may contribute to EDMs.  
In supersymmetric (SUSY) models, these parameters come from the soft 
SUSY breaking sector and the superpotential $\mu$-term and, if R-parity 
is not imposed, additional CPV phases appear from the R-parity violating 
trilinear and bilinear parameters. 

During the last few years significant progress has been achieved in 
experimental studies of various 
EDMs~\cite{neutron-exp}-\cite{deuteron-exp}.  
Presently there exist stringent upper bounds on the neutron EDM, 
$d_n$,~\cite{neutron-exp2} and the EDM, $d_{Hg}$, of the neutral 
$^{199}$Hg atom~\cite{Hg-exp}: 
\eq\label{exp-dn}  
|d_n| &\leq& 3.0 \times 10^{-26} \ e\cdot {\rm cm} \,,\\
\label{exp-Hg} 
|d_{Hg}| &\leq& 2.1 \times 10^{-28} \ e\cdot {\rm cm} \ .
\en 
Recently it was also proposed to measure the deuteron EDM, $d_D$, 
in storage ring experiments~\cite{deuteron-exp} with deuteron ions 
instead of neutral atoms. The advantage of these experiments is the  
absence of Schiff screening, which introduces significant uncertainties  
in the case of neutral atoms. This allows a direct probe of the value  
for $d_D$. In the near future it is hoped to obtain the experimental 
upper bound of 
\eq\label{exp-D} 
|d_D| \leq (1 \div 3) \times 10^{-27} \ e\cdot {\rm cm}\ . 
\en 
The upper limits for the EDMs, derived from the above null experimental 
results, stringently constrain or even reject various models of 
CPV~\cite{Khriplovic-Lamoreaux}. 
For the case of SUSY models with the superpartner masses around 
the electroweak scale $\sim$ 100 GeV $-$ 1 TeV, these limits imply 
that the CPV SUSY phases are very small. Various aspects of the 
calculation of the EDMs within the popular versions of SUSY models 
with~\cite{SUSY-EDM-1,Hisano} and 
without~\cite{BM86,SUSY-EDM-2,RPV-EDM,Herczeg-eN,Faessler:2006vi} 
R-parity conservation have been studied in the literature.

In the present paper we are studying the EDMs of the $^{199}$Hg atom and 
the deuteron as well as of the light baryons (nucleon and neutral 
hyperons $\Lambda$, $\Sigma^0$ and $\Xi^0$) in the framework of 
a generic SUSY model without R-parity conservation (\rp SUSY) on the 
basis of the SU(3) version of chiral perturbation 
theory (ChPT)~\cite{Hisano,ChPT,Pich:1991fq,Borasoy:2000pq}. 
We consider CP-violation in the hadronic 
sector originating from the quark chromoelectric dipole moments (CEDMs) 
and CPV 4-quark effective interactions which are induced by the complex 
phases of the trilinear \rp-couplings $\lambda^{\prime}$. 
From the experimental bounds of Eqs.~(\ref{exp-dn})-(\ref{exp-Hg})
we derive upper limits on the imaginary parts of the products of 
the trilinear \rp-couplings and compare these limits to the existing 
ones. On this basis we predict the values for the EDMs of the light 
neutral hyperons. We also  discuss the prospects of the deuteron 
EDM experiments (\ref{exp-D}) from the view point of their ability to 
improve these limits.

\section{Hadronic EDMs in Chiral Perturbation Theory} 
\label{s2}

Here we briefly outline the formalism we use for the calculation of 
the EDMs of the neutral $^{199}$Hg atom, the deuteron and light baryons 
$n$, $p$, $\Lambda$, $\Sigma^0$ and $\Xi^0$.  

The $^{199}$Hg is a diamagnetic atom with a closed electron shell. 
Its EDM is dominated by the nuclear CP-violating effects characterized 
by the Schiff moment $S_{Hg}$, generating a T-odd electrostatic potential 
for atomic electrons. The $^{199}$Hg atomic EDM is given by~\cite{Hg} 
\eq\label{HgEDM} 
d_{Hg}= - 2.8\times 10^{-4} S_{Hg} \cdot {\rm fm}^{-2}\,. 
\en 
The deuteron EDM is a theoretically rather clean problem~\cite{Khriplovich} 
since the deuteron represents the simplest nucleus with a well understood 
dynamics. The corresponding EDM can be written as the sum of the three terms
\eq\label{dEDM-0}                 
d_D= d_p + d_n + d_D^{NN}\,,
\en 
where $d_n$, $d_p$ are the neutron and proton EDMs, respectively, and 
$d_D^{NN}$ is due to the CP-violating nuclear forces.

For the evaluation of the proton-neutron CP-odd nuclear term, 
$d_D^{NN}$ and the Schiff moment $S_{Hg}$ we are using a one-meson 
($\pi$ or $\eta$) exchange model with CP-odd meson-nucleon 
interactions~\cite{Khriplovich-PhRep,Khriplovich,DS:2005,Hisano}. 

The baryon EDM $d_B$ is given by the value of the corresponding form 
factor at zero recoil, i.e. $d_B = D_B(0)$. The EDM form factor 
$D_B(Q^2)$ is defined in the standard way through the baryon matrix 
element of the electromagnetic current: 
\begin{eqnarray}\label{DB-def}
\langle B(p^\prime)| J_{\mu}(0) |B(p)\rangle &=&  
\bar{u}_n(p^{\prime}) \biggl[ \gamma_\mu \, F_B^1(Q^2)
\, + \, \frac{i}{2 m_B} \, \sigma_{\mu\nu} \, 
q^{\nu} \, F_B^2(Q^2) \\
&-& \sigma_{\mu\nu} \, \gamma_5 \, q^{\nu} \, \, D_B(Q^2)
\, + \, (\gamma_\mu \, q^2 \, - \, 2 \, m_N \, q_\mu) \, 
\gamma_5 \, A_B(Q^2)
\biggr] u_n(p) \,, \nonumber
\end{eqnarray}
where, in addition, $F_B^1(Q^2)$ and $F_B^2(Q^2)$ are the
well-known $CP$-even electromagnetic form factors and
$A_B(Q^2)$ is the baryon anapole moment form factor. 

In what follows we evaluate the EDMs of light baryons, $d_n$, $d_p$, 
$d_\Lambda$, $d_{\Sigma^0}$ and  
$d_{\Xi^0}$ on the basis of the SU(3) version of Chiral Perturbation 
Theory (ChPT)~\cite{Pich:1991fq,Borasoy:2000pq,Hisano}. 
We use the Lagrangian of SU(3) ChPT of order $O(p)$ and restrict 
ourselves to the meson-loop diagrams 
given in Fig.1. As it is known and was discussed before in the 
literature (see e.g. Ref.~\cite{Hisano}), an accurate calculation 
of the baryon EDMs should also include the contribution of the 
unknown low-energy constants (LECs) which parameterize the short-distance 
effects and remove the ultraviolet divergences. However, we assume that 
at the level of accuracy necessary for the analysis of the \rp SUSY in 
hadronic EDMs the one-loop meson-cloud 
approximation~\cite{Pich:1991fq,Borasoy:2000pq,Hisano} is adequate. 

The CP-conserving vertices of the diagrams in Fig.1 correspond to 
the terms of the ChPT Lagrangian which is given by the sum of leading 
meson and meson-baryon pieces:  
\eq\label{L_CHPT}  
{\cal L} \, = \, \frac{F_\pi^2}{4} \, \la{u_\mu u^\mu + \chi_+}\ra \, + \, 
\la \bar B \, (i \not\!\! D - m_B ) \, B \ra 
\, + \, \frac{D}{2} \la \bar B \gamma^\mu \gamma^5 \{u_\mu B \} \ra 
\, + \, \frac{F}{2} \la \bar B \gamma^\mu \gamma^5 [u_\mu B ] \ra  \, 
\nonumber 
\en 
where $D=0.80$ and $F=0.46$ are the baryon axial coupling constants, 
$m_B$ is the baryon mass in the chiral limit, the symbols 
$\left\langle \ldots \right\rangle $, $\{ \ldots \}$ and 
$[ \ldots ]$ denote the trace over flavor matrices, 
anticommutator and commutator, respectively. 

We use the standard notation for the basic blocks of the ChPT 
Lagrangian~\cite{ChPT} where
\eq 
U=u^{2}=\exp (iP\sqrt{2}/F_\pi)
\en 
is the chiral field collecting pseudoscalar fields ${\cal P}$ 
in the exponential pa\-ra\-me\-tri\-za\-tion with 
$F_\pi=92.4$ MeV being the octet leptonic decay constant, 
$D_{\mu }$ denotes the chiral and gauge-invariant 
derivative, $u_{\mu }=iu^{\dagger }D_{\mu}Uu^{\dagger }$, 
\hspace*{0.2cm} $\chi _{\pm }=u^{\dagger }\chi u^{\dagger
}\pm u\chi ^{\dagger }u,\hspace*{0.2cm}\chi =2B(s+ip),\,\,\,s={\cal M} 
+\ldots \,$ and ${\cal M}={\rm diag}\{m_u,m_d,m_{s}\}$ 
are the charge and the mass matrix of current quarks, respectively;   
$B$ is the quark vacuum condensate parameter. 
The explicit form of the octet matrices of pseudoscalar mesons $P$ 
and baryons $B$ can be found e.g. in 
Refs.~\cite{Pich:1991fq,Borasoy:2000pq,Hisano}.  
In our analysis we take into account $\pi^0 - \eta$ meson 
mixing~\cite{ChPT} with the corresponding mixing angle $\varepsilon$ given 
by 
\eq 
{\rm tan} 2\varepsilon \, = \, 
\frac{\sqrt{3}}{2} \, \frac{m_d - m_u}{m_s - \hat m} 
\en 
where $\hat m = (m_u + m_d)/2$. In the numerical calculations we use 
the standard set of current quark masses: 
$m_u = 5$~MeV, $m_d = 9$~MeV and $m_s = 175$~MeV. For  
the pion and kaon masses we use the values of the charged mesons: 
$M_\pi = M_{\pi^\pm} = 139.57$~MeV and $M_K = M_{K^\pm} = 493.677$~MeV. 
For the baryon masses 
we use the universal parameter, the value of the octet baryon mass 
in the chiral limit, which for convenience is identified with the 
proton mass: $m_B = m_p = 938.27$~MeV. 
Note that the Lagrangian (\ref{L_CHPT}) 
generates the CP-even meson-baryon, photon-meson and photon-baryon 
coupling. 

The CP-odd meson-baryon Lagrangian has been derived in Ref.~\cite{Hisano}
where one can find its complete form. 
Here we explicitly only show the CPV pion-nucleon terms: 
\eq\label{piNN_CPV} 
\hspace*{-.75cm} 
{\cal L}^{CPV}_{M BB} &=& 
\bar{g}_{M B B} \bar B M B  \, = \, 
\overline{N} \, \biggl\{ \, 
\bar{g}^{(0)}_{\pi NN} \, \vec{\pi} \, \vec{\tau} \, 
+ \, \bar{g}^{(1)}_{\pi NN} \, \pi^0  \, 
+ \, \bar{g}^{(2)}_{\pi NN} \, ( \vec{\pi} \, \vec{\tau} \, 
- \, 3 \, \pi^0 \, \tau^3) \, \biggr\} \, N  + ... 
\en 
where $\bar{g}^{(0)}_{\pi NN}$, $\bar{g}^{(1)}_{\pi NN}$ and 
$\bar{g}^{(2)}_{\pi NN}$ are the corresponding isoscalar, isovector 
and isotensor coupling constants. 

\newpage

\section{EDMs in \rpp SUSY: quark level}
\label{s3}

The effective CP-odd Lagrangian in terms of quark, gluon and photon 
fields up to operators of dimension~six, normalized at the hadronic scale 
$\sim$1 GeV, has the following standard form: 
\eq\label{CPV-q} 
{\cal L}^{CPV} &=& \frac{\bar\theta}{16 \pi^{2}} {\rm tr}
\big(\widetilde{G}_{\mu\nu} G^{\mu\nu} \big) 
- \frac{i}{2} \sum_{i = u,d,s} d_i \ \bar q_i \, 
\sigma^{\mu \nu} \, \gamma_5 \, e F_{\mu\nu} \, q_i \\
&-& \frac{i}{2} \sum_{i = u,d,s} {\tilde d}_i\  \bar q_i \, 
\sigma^{\mu \nu} \, \gamma_5 
\, g_s G_{\mu\nu}^a \, T^a \, q_i \, 
\, - \,  \frac{1}{6} \, C_W \,
f^{abc} \, G^a_{\mu\alpha}  \, G^{b\alpha}_{\nu}
\, G^c_{\rho\sigma} \, \varepsilon^{\mu\nu\rho\sigma} \,, \nonumber 
\en
where $G^a_{\mu\nu}$ is the gluon stress tensor,
$\widetilde{G}_{\mu\nu}=\frac{1}{2}
\epsilon_{\mu\nu\sigma\rho}{G}^{\sigma \rho}$ is its dual tensor,
and  $T^a$ and $f^{abc}$ are the $SU(3)$ generators and structure
constants, respectively. In this equation the first term
represents the SM QCD $\theta$-term, while the last three terms
are the non-renormalizable effective operators induced by physics
beyond the SM. The second and third terms are the dimension-five
electric and chromoelectric dipole quark operators, respectively,
and the last term is the dimension-six Weinberg operator. 
The light quark EDMs and CEDMs are denoted by $d_i$ and $\tilde{d}_i$,
respectively. 
In what follows we adopt the Peccei-Quinn mechanism, eliminating 
the $\bar\theta$-term as an independent source of CPV. 

We also consider the 4-quark 
CPV interactions of the form~\cite{4-fermion-1,4-fermion-2} 
\eq\label{4-q} 
{\cal L}^{CPV}_{4q} = \sum_{i,j} \bigg\{
C^P_{ij}(\bar{q}_i q_i)(\bar{q}_j i \gamma_5 q_j) +
C^T_{ij}(\bar{q}_i \sigma_{\mu\nu} q_i)
(\bar{q}_j i \sigma^{\mu\nu} \gamma_5 q_j)\biggr\}\,,
\en 
where the sum runs over all the quark flavors $i,j=u,d,s,c,b,t$.
The operators of the above Lagrangians in Eqs.~(\ref{CPV-q})-(\ref{4-q}) 
can be induced by physics beyond the SM at loop or tree level after 
integrating out the heavy degrees of freedom.

Here we are studying the CPV effects in the hadronic sector induced by the 
trilinear interactions of \rp SUSY models. The corresponding part 
of the $R_p$-violating superpotential reads: 
\eq\label{W_rp}
W_{R_p \hspace{-0.8em}/\;\:} = \lambda^{\prime}_{ijk}L_i Q_j D^c_k \,, 
\end{eqnarray} 
where the summation over the generation indices $i, j, k$ is understood, 
$L$, $Q$ and $D^c$ are the superfields of lepton-sleptons, 
quarks-squarks and $CP$-conjugated quarks-squarks, respectively,   
and $\lambda^{\prime}_{ijk}$ 
are the complex coupling constants violating lepton number conservation. 
Eq.~(\ref{W_rp}) results in the interactions
\eq\label{lambda}
{\cal L}_{\lambda^\prime} \, = \, 
- \, \lambda _{ijk}^{\prime} \, ( \, 
\tilde{\nu}_{_{iL}}\bar{d}_{_k} P_{L} d_{_j} +
\tilde{d}_{_{jL}}\bar{d}_{_k} P_{L} \nu _{_i} +
\tilde{d}_{_{kR}}\bar{d}_{_j} P_{R} \nu^c_{_i}  
\, - \,  \tilde{l}_{_{iL}}\bar{d}_{_k} P_{L} u_{_j}   
- \tilde{u}_{_{jL}}\bar{d}_{_k} P_{L} l_{_i}-
\tilde{d}_{_{kR}} \bar{u}_{_j} P_{R} l^c_{_i} \, ) \ +\ \mbox{H.c.} 
\en 
with $P_{L,R} = (1 \mp \gamma_5)/2$. 

The interactions of the Lagrangian (\ref{lambda}) generate the terms in the 
effective CPV Lagrangians~(\ref{CPV-q}) and (\ref{4-q}) at certain orders 
in the $\lambda^{\prime}$-couplings.  
It is straightforward to derive the corresponding contribution to the 
4-quark contact terms~(\ref{4-q}). It arises from a tree level contribution 
induced by sneutrino ($\tilde\nu$) exchange given by 
\eq\label{4-q-RPV} 
{\cal L}^{CPV}_{4q} = [C^P_{sd} (\bar{s} s) 
                     + C^P_{bd} (\bar{b} b)] 
\ (\bar{d} i \gamma_5 d) \,
\en
with
\eq\label{C-P} 
C^P_{sd} = \sum\limits_{i} \frac{{\rm Im} (\lambda^{\prime}_{i22} 
\lambda^{\prime *}_{i11})}{2 m^2_{\tilde\nu(i)}}, \ \ \ \ \ 
C^P_{bd} = \sum\limits_{i} \frac{{\rm Im} (\lambda^{\prime}_{i33} 
\lambda^{\prime *}_{i11})}{2 m^2_{\tilde\nu(i)}}\,,
\en 
where $m_{\tilde\nu}$ is the sneutrino mass. Note, that the four-quark 
term involving only $d$-quarks is absent in (\ref{4-q-RPV}) due to 
${\rm Im} (\lambda^{\prime}_{i11} \lambda^{\prime *}_{i11}) \equiv 0$. 

The interactions of the Lagrangian~(\ref{lambda}) generate the quark EDMs, 
$d_q$, and CEDMs, $\tilde{d}_q$, starting from 
2-loops~\cite{SUSY-EDM-2,RPV-EDM} and the dominant \rp-contributions 
are of second order in the $\lambda^{\prime}$-couplings. 
It was shown in Ref. \cite{RPV-EDM} that the up-quark EDM and CEDM are 
suppressed by the light quark mass and mixing angles, which, therefore,   
can be neglected. The quark EDMs are irrelevant for our study based on 
the pion-exchange model with the interaction Lagrangian~(\ref{piNN_CPV}). 
We also do not consider the Weinberg term, which does not appear 
at the order of $O(\lambda^{\prime \, 2})$ unlike the quark CEDMs and 
4-quark contact terms. In our analysis we use for the d-quark CEDMs 
the 2-loop result of Ref.~\cite{RPV-EDM}. The dominant contribution coming 
from the virtual b-quark takes the form: 
\eq 
\tilde d_k &=& 6.2 \times 10^{-7} ({\rm GeV}^{-1}) \, \sum\limits_{i} 
{\rm Im} (\lambda^{\prime}_{i33} \lambda^{\prime\ast}_{i11}) 
\ \, {\cal F}\left(\frac{m_b^2}{m^2_{\tilde\nu(i)}}\right), 
\en 
where $k=1,2$ and $d_1\equiv d_d,\  d_2\equiv d_s$.  
The scaling factor ${\cal F}$ originates from the loop 
integration and can be written as
\begin{eqnarray}\label{scale-1}
{\cal F}(\tau) = \frac{F(\tau)}{F(\tau_{300})}\,, \hspace*{.5cm} 
F(\tau) = \tau\left[\frac{\pi^2}{3} +2 + \ln\tau + 
\left(\ln\tau\right)^2\right],
\end{eqnarray}
where $\tau_{300} = (m_b/300 \, {\rm GeV})^2$ and $m_b=4.5$ GeV. 
For convenience the scaling factor ${\cal F}$ is normalized as 
${\cal F}(\tau_{300}) = 1$ which corresponds to $m_{\tilde\nu}=300$ GeV.

\section{EDMs in \rpp SUSY: hadronic level}
\label{s4}

In order to link the CP-violation at the quark and hadronic levels  
we have to relate the parameters of the Lagrangian in Eq. (\ref{piNN_CPV}) 
to the quark CEDMs, $\tilde d_q$, and the CPV 4q-couplings $C^P_{ij}$.
Towards this end we apply the standard matching of the 
quark-level~(\ref{CPV-q})-(\ref{4-q}) and hadronic-level (\ref{piNN_CPV}) 
Lagrangians. This allows us to express the CP-odd meson-baryon couplings 
in terms of the quark CEDMs~\cite{Hisano} and  the
CPV 4q-couplings $C^P_{ij}$ as 
\eq \label{gg0}
\bar{g}^{(0)}_{\pi NN} &=& \frac{\la \bar u u - \bar d d \ra}{2F_\pi} \, 
\biggl\{ A_u + A_d - \frac{\varepsilon}{3\sqrt{3}}(A_u - A_d) \biggr\}\,, \\
\bar{g}^{(1)}_{\pi NN} &=& \frac{\la \bar u u + \bar d d \ra}{2F_\pi} \, 
\biggl\{ A_u - A_d - \frac{\varepsilon}{\sqrt{3}}(A_u + A_d) \biggr\} 
\,  + \, \frac{\la \bar s s \ra}{2F_\pi}  
\, \frac{4\varepsilon}{\sqrt{3}} \, A_s  \nonumber\\
&-& F_{\pi} \frac{M_{\pi}^2}{2m_d}\left(C^P_{sd} \la \bar s s \ra + 
C^P_{bd} \la \bar b b \ra \right)\,, \\
\label{gg2}
\bar{g}^{(2)}_{\pi NN} &=& \frac{\la \bar u u - \bar d d \ra}{2F_\pi} 
\,  \frac{\varepsilon}{3\sqrt{3}} \, (A_u - A_d), \ \  
\bar g_{K^+ n \Sigma^-} =  
\frac{\la \bar s s - \bar d d \ra}{2F_\pi} \, (A_u + A_s) \,, 
\ \ \ \ \cdots 
\en  
If the Peccei-Quinn (PQ) symmetry is imposed the parameters
$A_q$ are expressed through the quark CEDMs $\tilde d_q$  
as $A_q = - 0.27 \tilde d_q$ GeV. Here $\la \bar q q \ra \equiv 
\la p | \bar q q | p \ra$ are the scalar quark condensates in the 
proton~\cite{Hisano,Gsb1,Gsb2}: 
\eq 
\la \bar u u \ra = 3.5\,, \hspace*{.25cm} \la \bar d d \ra = 2.8\,, 
\hspace*{.25cm} \la \bar s s \ra = (0.64 \div 3.9) \,, 
\hspace*{.25cm} \la \bar b b \ra = 9\times 10^{-3} \,. 
\en 
Note that the values of strange and bottom condensates in the nucleon 
are subject to significant uncertainties. In our analysis we use 
the estimates from Refs.~\cite{Gsb1,Gsb2}. 
For the value of $\la \bar s s \ra $ we indicate the interval 
according to Ref.~\cite{Gsb1}. For $\la \bar b b \ra $ we only need 
an order of magnitude estimate since it is associated 
with the subdominant term not essential for our analysis. 

Now we are in the position to calculate the 
diagrams which contribute to the baryon EDMs (see Fig.1). 
The calculation of the Feynman diagrams in Fig.1 is 
straightforward and discussed before e.g. 
in Refs.~\cite{Pich:1991fq,Borasoy:2000pq}. The diagrams in Fig.1a and 1b 
contribute to the chiral logarithms~\cite{pnEDM-Chiral}, 
the constant terms plus higher-order terms which can be neglected 
in the chiral expansion: $\sim [ {\rm log}(m_B/M_P) - 1 + O(M_P) ]$.  
The diagrams in Fig.1c and 1d are dominated by the constant terms in 
the chiral expansion: $\sim [ 1 + O(M_{P}) ]$. 
Finally, the diagrams in Fig.1e and 1f cancel 
each other. For the neutral baryons ($n$, $\Lambda$, $\Sigma^0$ and 
$\Xi^0$) both sets of diagrams in Fig.1a,b and Fig.1c,d contribute 
in such a way that the constant terms cancel each other. This is not the 
case for the EDMs of charged baryons where both chiral logarithms and 
constant terms give a non-trivial contribution. In the case of the 
neutron EDM the leading-order contributions of the diagrams in Figs.1a,b 
and Figs.1c,d in terms of CP-even and CP-odd meson-baryon coupling 
constants  are~\cite{Hisano}: 
\eq
d_n^{1(a+b)} = \frac{e g_{\pi NN} \bar g_{\pi NN}}{4\pi^2 m_B} \, 
\biggl[ {\rm log}\frac{m_B}{M_\pi} - 1 \biggr]  \, - \, 
\frac{e g_{K N \Sigma} \bar g_{K^+ n \Sigma^-}}{4\pi^2 m_B} \, 
\biggl[ {\rm log}\frac{m_B}{M_K} - 1 \biggr]  
\en 
and 
\eq
d_n^{1(c+d)} = \frac{e g_{\pi NN} \bar g_{\pi NN}}{4\pi^2 m_B} \,  
\, - \, 
\frac{e g_{K N \Sigma} \bar g_{K^+ n \Sigma^-}}{4\pi^2 m_B} \,,  
\en 
where 
\eq 
& &g_{\pi NN} = \frac{m_B}{F_\pi} \, (D + F)\,, \quad\quad 
g_{K N\Sigma} = \frac{m_B}{F_\pi} \, (D - F)\,,\\ 
& &\bar g_{\pi NN} = \bar g_{\pi NN}^{(0)} + \bar g_{\pi NN}^{(2)}\,. 
\en 
The complete result for the neutron EDM is: 
\eq\label{d_n}
d_n = \frac{e g_{\pi NN} \bar g_{\pi NN}}{4\pi^2 m_B} \, 
{\rm log}\frac{m_B}{M_\pi}   \, + \, 
\frac{e g_{K N \Sigma} \bar g_{K^+ n \Sigma^-}}{4\pi^2 m_B} \, 
{\rm log}\frac{m_B}{M_K}. 
\en 
Substituting the expressions of the baryon-meson couplings 
Eqs.~(\ref{gg0})-(\ref{gg2}) in terms of the parameters of the 
ChPT Lagrangian and quark CEDMs~\cite{Hisano} we have 
\eq \label{Dn}
\hspace*{-.5cm} 
d_n = \beta \, \tilde d_{ud}^+ \, c_{ud}^- \, (D+F) \, 
{\rm log}\frac{m_B}{M_\pi} 
+  \beta \, \tilde d_{us}^+ \, c_{ds}^- \, (D-F) \, 
{\rm log}\frac{m_B}{M_K} \, 
\en 
where $\tilde d_{q_1q_2}^\pm = \tilde d_{q_1} \, \pm \, 
\tilde d_{q_2}\,,$  
$c_{q_1q_2}^\pm =  \la \bar q_1 q_1 \pm \bar q_2 q_2 \ra$ and 
$\beta = 0.27e/(8\pi^2 F_\pi^2)$. In Eq.~(\ref{d_n}) we neglected 
the pion-eta meson mixing ($\varepsilon = 0$). 

In a similar way we calculate the EDMs of other baryons. 
Here we indicate the final results of these calculations: 
\eq \label{Dpp}
d_p &=& - \, \beta \, \tilde d_{ud}^+  \, c_{ud}^- \, 
(D+F) \, {\rm log}\frac{m_B}{M_\pi} \, - \, 
\beta \, \tilde d_{us}^+ \, [ \, F \, c_{us}^- 
\, + \, \frac{D}{3} \, (c_{ud}^-  - c_{ds}^-)\, ] 
{\rm log}\frac{m_B}{M_K} \nonumber\\
&+&\frac{\beta}{3} \tilde d_u [ D(5 c_{ud}^- + c_{us}^+) 
+ 3 F \, ( 2c_{ud}^- + 2c_{us}^- + c_{ds}^+) \, ]  \nonumber \\ 
&+&\frac{\beta}{3} \tilde d_d [ D \, ( 5 c_{ud}^- - 5 c_{ds}^- 
- 2 c_{us}^+ ) 
+ 3 F \, c_{us}^-  ]  \nonumber\\
&+&\frac{\beta}{3} \tilde d_s [ D \, (c_{us}^+ - 5 c_{ds}^-)  
\, + \, 3 F \, (2c_{us}^- + 2c_{ds}^- -  c_{ud}^+) \, ]  \,, \\[2mm] 
d_\Lambda &=& - d_{\Sigma^0} \ = \  
\frac{\beta}{2} \, \tilde d_{us}^+  \, 
[ \, D \, c_{us}^- \, + \, F (c_{ud}^- - c_{ds}^-) \, ] \, 
{\rm log}\frac{m_B}{M_K} \,,\\
\label{Xi0}
d_{\Xi^0} &=& \beta \, \tilde d_{ud}^+  \, 
c_{ds}^- \, (D - F) \, {\rm log}\frac{m_B}{M_\pi} \, + \, 
\beta \, \tilde d_{us}^+  \, 
c_{ud}^- \, (D + F) \, {\rm log}\frac{m_B}{M_K} \,. 
\en 
As it is known~\cite{Hisano} the proton and neutron contributions 
to the deuteron EDM cancel out in leading order of the chiral 
expansion in the SU(2) version of ChPT~\cite{EDMd-1}. 
However it does not hold in the SU(3) extension~\cite{Hisano}. 
Therefore, the contribution of the strangeness sector to the deuteron 
EDM becomes important. Note, in the final expressions for the neutron  
and proton EDMs, Eqs. (\ref{Dn}), (\ref{Dn}), we disagree with the  
results of Ref.~\cite{Hisano} by a factor 2.  
We discuss this issue in Appendix.  

The deuteron EDM also receives a contribution from  
P- and T-odd proton-neutron forces generated 
by $\pi$- and $\eta$-meson  exchange between  
two nucleons with one CP-even and one 
CP-odd vertex. With the corresponding potential 
one can calculate the $d_D^{NN}$ term in Eq.~(\ref{dEDM-0}) as 
the expectation value of $e {\bf r}/2$, where ${\bf r}$ is the 
relative proton-neutron coordinate. In this way one obtains the 
following result~\cite{Khriplovich,Khriplovich-PhRep}:
\eq\label{dEDM-1}
d_D^{NN}\,=\,-\,{e \, g_{\pi NN} \, \bar{g}_{\pi NN}^{(1)} 
\over 12 \pi m_{\pi}} \, {1+\xi\over (1+2\xi)^2} \,,
\en
where $\xi = \sqrt{M_N E_B}/m_{\pi}$ and $E_B=2.23$ MeV is 
the deuteron binding energy. Expressing the CP-even and CP-odd 
pion-nucleon couplings in terms of the CEDMs and parameters of 
the chiral Lagrangian we get: 

\eq\label{dEDM-2}
d_D^{NN} \simeq  \,    
\gamma_{D}^{(1)} \, [ \,  \tilde d_{ud}^- \, c_{ud}^+ 
\, + \, \frac{4 \varepsilon}{\sqrt{3}} 
\tilde d_s \la \bar s s \ra \, ]  
-\gamma_{D}^{(2)} \, [ \,  C^P_{sd} \la \bar s s \ra + 
C^P_{bd} \la \bar b b \ra \, ] 
\en 
where 
\eq 
\gamma_D^{(1)} \, = \, 0.13 \, (F+D) \, 
\frac{e m_B}{24\pi M_\pi F_\pi^2}\,, \hspace*{.5cm} 
\gamma_D^{(2)} \, = \, 0.48 \, (F+D) \, 
\frac{e m_B M_\pi}{24\pi m_d}\,. 
\en 
Recently, the Schiff moment $S_{Hg}$~\cite{DS:2005} has been 
calculated within a reliable nuclear structure model which 
takes full account of core polarization on the basis of 
the P- and T-odd one-pion exchange potential. 
Note that in Ref.~\cite{Hisano} it was shown that 
the contribution of the eta-meson exchange to the Schiff moment 
is suppressed by a factor $10^{-3}$ with respect to the pion 
exchange. The result for the Schiff moment, taking into 
account a finite range interaction and the core polarization 
effect is 
\eq\label{Schiff-1}
S_{Hg} \, = \, - \, 0.055 \, g_{\pi NN} \, \{ 
0.007 \bar g_{\pi NN}^{(0)}  + \bar g_{\pi NN}^{(1)} 
- 0.16 \bar g_{\pi NN}^{(2)} \}  \ e \cdot {\rm fm}^3 \,.
\en 
Therefore, only the isovector channel is sufficient for the 
evaluation of the Schiff moment~\cite{DS:2005,Hisano}. 
In terms of the quark EDMs and ChPT parameters the latter 
is given by  
\eq\label{c}
S_{Hg} \simeq  \, \biggl\{  
\gamma_{Hg}^{(1)} \, [ \,  \tilde d_{ud}^- \, c_{ud}^+ 
\, + \, \frac{4 \varepsilon}{\sqrt{3}} 
\tilde d_s \la \bar s s \ra \, ] 
- 
\gamma_{Hg}^{(2)} \, [ \,  C^P_{sd} \la \bar s s \ra + 
C^P_{bd} \la \bar b b \ra \, ]  \biggr\} 
\ e \cdot {\rm fm}^3 \,,\ \ 
\en 
where 
\eq 
\gamma_{Hg}^{(1)} \, = \, 0.015 \, (F+D) \, \frac{m_B}{2 F_\pi^2}\,,  
\hspace*{.5cm} 
\gamma_{Hg}^{(2)} \, = \, 0.055 \, (F+D) \, \frac{m_B M_\pi^2}{2 m_d} \,. 
\en

\section{Constraints on \rpp SUSY from hadronic EDMs}
\label{s5}

Let us summarize the formulas for the considered hadronic EDMs in terms 
of the trilinear \rp-couplings. Using Eqs.~(\ref{HgEDM}), (\ref{dEDM-0}), 
(\ref{Dn})-(\ref{Xi0}), (\ref{dEDM-2}) and (\ref{c}) we get 
the following expressions with numerical coefficients:  
\eq\label{num1}  
d_p &=& -  10^{-20}\times {\cal F}
\left(\frac{m_b^2}{m^2_{\tilde\nu(i)}}\right)\left[(1.67 \div 2.21) \, 
\,  {\rm Im} (\lambda^{\prime}_{i33} \lambda^{\prime\ast}_{i11}) 
+ (0.23 \div 0.48) \, 
{\rm Im} (\lambda^{\prime}_{i33} \lambda^{\prime\ast}_{i22})\right] 
\ e \cdot {\rm cm} \,, \\[3mm] 
\label{num2}  
d_n &=&  10^{-20}\times {\cal F}\left(\frac{m_b^2}{m^2_{\tilde\nu(i)}}\right)
\left[0.82 \, 
{\rm Im} (\lambda^{\prime}_{i33} \lambda^{\prime\ast}_{i11})
+ (-0.12 \div 0.23) \, 
{\rm Im} (\lambda^{\prime}_{i33} \lambda^{\prime\ast}_{i22})\right] 
\ e \cdot {\rm cm} \,, \\[3mm] 
\label{num3}  
d_{Hg} &=&  10^{-23}\times 
{\cal F}\left(\frac{m_b^2}{m^2_{\tilde\nu(i)}}\right)
\left[11.4 \, {\rm Im} (\lambda^{\prime}_{i33} \lambda^{\prime\ast}_{i11}) 
+ (0.28 \div 1.62) \, 
{\rm Im} (\lambda^{\prime}_{i33} \lambda^{\prime\ast}_{i22})\right] 
\ e \cdot {\rm cm}
\nonumber\\
&-& (0.90 \div 5.49) \times 10^{-23}  
\left(\frac{300 \, {\rm GeV}}{m_{\tilde\nu(i)}}\right)^2 \, 
{\rm Im} (\lambda^{\prime}_{i22} \lambda^{\prime\ast}_{i11}) 
\ e \cdot {\rm cm}\,, \\[3mm] 
\label{num4}  
d_D &=&  10^{-20}\times 
{\cal F}\left(\frac{m_b^2}{m^2_{\tilde\nu(i)}}\right)
\left[(11.79 \div 12.34) \, 
{\rm Im} (\lambda^{\prime}_{i33} \lambda^{\prime\ast}_{i11}) 
+ (-0.41 \div 0.03) \, {\rm Im} (\lambda^{\prime}_{i33} 
\lambda^{\prime\ast}_{i22})\right] 
\ e \cdot {\rm cm} \, 
\nonumber\\
&-& (0.4 \div 2.5) \times 10^{-20}\left(\frac{300 \, 
{\rm GeV}}{m_{\tilde\nu(i)}}\right)^2 \, 
{\rm Im} (\lambda^{\prime}_{i22} \lambda^{\prime\ast}_{i11}) 
\ e \cdot {\rm cm} \,, \\[3mm] 
\label{num5}  
d_\Lambda &=& - \, d_{\Sigma^0} \ = \ (0.08 \div 0.25) 
\times 10^{-20} {\cal F}\left(\frac{m_b^2}{m^2_{\tilde\nu(i)}}\right)\, 
{\rm Im} (\lambda^{\prime}_{i33} \lambda^{\prime\ast}_{i22}) \ 
e \cdot {\rm cm} \,, \\[3mm]
\label{num6}  
d_{\Xi^0} &=& 10^{-20}  {\cal F}\left(\frac{m_b^2}{m^2_{\tilde\nu(i)}}\right)
\left[(-0.35 \div 0.69) \, 
{\rm Im} (\lambda^{\prime}_{i33} \lambda^{\prime\ast}_{i11}) 
+ 0.28 \, 
{\rm Im} (\lambda^{\prime}_{i33} \lambda^{\prime\ast}_{i22})\right] 
\ e \cdot {\rm cm}.
\en 
In the above equations the summation over $i=1,2,3$ is implied.
The uncertainties in the coefficients are due to the variation 
of the strange quark sea in the proton.  
Note that the contribution from 
${\rm Im} (\lambda^{\prime}_{i22} \lambda^{\prime\ast}_{i11})$
appears solely via the 4-quark CPV interactions~(\ref{4-q-RPV}).

Comparing Eqs.~(\ref{num2}) and (\ref{num3}) with the corresponding 
experimental bounds Eqs.~(\ref{exp-dn}) and (\ref{exp-Hg}) we derive 
constraints on the imaginary parts of the products of \rp-couplings 
given in Table I. In the last column of Table I we also show for 
comparison the existing limits on $|\lambda^{\prime}_{i33} 
\lambda^{\prime\ast}_{i11}|$, $|\lambda^{\prime}_{i22} 
\lambda^{\prime\ast}_{i11}|$ and $|\lambda^{\prime}_{i33} 
\lambda^{\prime\ast}_{i22}|$~\cite{RPV-rev}. 
It is seen that the presently most stringent limits on 
$|{\rm Im}(\lambda^{\prime}_{ikk} \lambda^{\prime\ast}_{i11})|$, 
$|{\rm Im} (\lambda^{\prime}_{i33} \lambda^{\prime\ast}_{i22})|$
come from the $^{199}$Hg atom EDM (\ref{exp-Hg}). The forthcoming 
experiments on the deuteron EDM~(\ref{exp-D}) are going to improve 
these limits by about one to three orders of magnitude. 
Note, that we obtained about 1-order of magnitude improvement for the limit 
$|{\rm Im} (\lambda^{\prime}_{i33} 
\lambda^{\prime\ast}_{i11})|\leq 1.2\times 10^{-5}$ previously derived
in Ref. \cite{RPV-EDM} from the neutron EDM constraint~(\ref{exp-dn}) 
on the basis of the SU(6) quark model. 
The existing limits on the absolute values of the corresponding products 
do not exclude the values of 
$|{\rm Im} (\lambda^{\prime}_{ikk} \lambda^{\prime\ast}_{i11})|$, 
$|{\rm Im} (\lambda^{\prime}_{i33} \lambda^{\prime\ast}_{i22})|$ 
within the limits derived from EDMs. 
Using the limits from Table I we can predict on the basis of 
Eqs.~(\ref{num5}), (\ref{num6}) for the EDMs of neutral light hyperons 
the following upper limits: 
\begin{eqnarray}
|d_\Lambda| =   |d_{\Sigma^0}| \leq 1.9 \times 10^{-25} \ 
e \cdot {\rm cm}, \ \ \ \ 
|d_{\Xi^0}| \leq 2.4 \times 10^{-25} \ e \cdot {\rm cm} \, \nonumber
\end{eqnarray}
which might have some future phenomenological implications. 

\section{Summary}
\label{s6}

We have studied the contributions of the trilinear \rp-couplings 
to the EDMs of $^{199}$Hg atom,  deuteron, nucleon and neutral 
hyperons within the SU(3) ChPT, applying the meson-exchange model 
of CPV nuclear forces. We have analyzed the \rp-contributions via 
the d-quark CEDM and CPV 4-quark interactions.
We have shown that the latter contribute only to the nuclear EDMs 
via the CPV nuclear forces and are irrelevant for the EDMs of the nucleon 
and neutral hyperons. We have also found that these two mechanisms give 
rise to a dependence of the hadronic EDMs 
proportional to different $\lambda^{\prime}$-couplings.
Therefore, taking into account both mechanism allows one 
to obtain a complimentary information on the imaginary parts of 
the products of the $\lambda^{\prime}$-couplings. The corresponding 
upper limits from the null experimental results on measurements of 
the above mentioned hadronic EDMs are given in Table 1.
On the basis of the derived constraints on the trilinear \rp-couplings 
we have given predictions for the EDMs of neutral hyperons which might 
have some phenomenological implications in future.

We have also demonstrated that the present limits from the $^{199}$Hg EDM 
experiments are by a factor $\sim$6 more stringent than those from the 
experiments on the neutron EDM and that the planned storage ring experiments 
with the deuterium ions would be able to significantly improve these limits.

\section{Acknowledgments}

This work was supported in part by the FONDECYT project 1030244,
by the DFG under the contracts FA67/31-1 and GRK683.
This research is also part of the EU Integrated Infrastructure
Initiative Hadronphysics project under the contract number
RII3-CT-2004-506078 and President grant of Russia "Scientific
Schools"  No 5103.2006.2. 

\appendix\section{On neutron and proton EDMs}

As we mentioned in Sec.IV our final expressions, Eqs.~(\ref{Dn}), 
(\ref{Dpp}), for the neutron and proton EDMs disagree with the results 
of Ref. [7] by a factor of 2. In order to check these results we compare 
them with the well known result of chiral perturbation theory 
in the two-flavor scheme, involving only pion loops. In this case all 
chiral approaches 
(see Refs.~\cite{Pich:1991fq,Borasoy:2000pq} and~\cite{pnEDM-Chiral}) 
give the same model-independent 
expression for the leading order neutron and proton EDMs in 
the chiral expansion, the so-called ``chiral logarithm". 
Neglecting kaon loops in our formulas Eqs. (\ref{Dn}), (\ref{Dpp}) 
we reproduce the result of the chiral approaches: 
\eq\label{A1} 
d_n \, = \, - d_p \, = \, 
\frac{e \, g_{\pi NN} \, \bar g_{\pi NN}}{4 \, \pi^2 \, m_p} 
\, {\rm log}\frac{m_p}{M_\pi} \, . 
\en
On the contrary, the result of Ref.~\cite{Hisano} given in their Eq.(50)
\eq \label{Hisano1}
d_n = - d_p = \frac{e}{4 \, \pi^2 \, F_\pi^2} \, (D + F) \, 
(A_u + A_d) \, ( \la \bar u u \ra  \, -  \, \la \bar d d \ra ) \, 
{\rm log}\frac{m_p}{M_\pi} 
\en 
differs from this formula by a factor 2. 
Indeed, using the expressions for $g_{\pi NN}$ and $\bar g_{\pi NN}$: 
\eq 
g_{\pi NN} \, = \, (D + F) \, \frac{m_p}{F_\pi}\,, \hspace*{.5cm}
\bar g_{\pi NN} = (A_u + A_d) \, 
\frac{\la \bar u u \ra  \, -  \, \la \bar d d \ra}{2 \, F_\pi} \, 
\en 
one can rewrite Eq. (\ref{Hisano1}) in the form  
\eq 
d_n \, = \, - d_p \, = \, \frac{e \, g_{\pi NN} \, 
\bar g_{\pi NN}}{2 \, \pi^2 \, m_p} \, {\rm log}\frac{m_p}{M_\pi} \, , 
\en 
which disagrees with Eq.~(\ref{A1}) by the factor 2 in the denominator.

%\newpage

\vspace*{.5cm} 
\begin{table}
\label{Table-1}
\caption{
Upper limits on the imaginary parts of the products of the trilinear 
\rp-couplings derived from the experimental bounds on the EDMs of 
the neutron \protect\cite{neutron-exp},  
the neutral $^{199}$Hg atom \protect\cite{Hg-exp} 
and the deuteron \protect\cite{deuteron-exp}. 
The existing constraints from other experiments on the absolute values 
of the corresponding products of \rp-coupling are taken from 
Ref.~\protect\cite{RPV-rev}. 
The scaling factor ${\cal F}$ is defined in Eq.~(\ref{scale-1}) 
and takes the values ${\cal F}$ = 1, 0.34 and 0.15 for 
$m_{\tilde\nu}$ = 300 GeV, 600 GeV and 1 TeV, respectively.} 

\vspace*{.2cm} 

\def\arraystretch{1.}
\begin{center}
\begin{tabular}{|c|c|c|c|c|}
\hline
Couplings& $d_n$ \protect\cite{neutron-exp} 
          & $d_{Hg}$ \protect\cite{Hg-exp} 
          & $d_D$ \protect\cite{deuteron-exp}
          & Existing limits \protect\cite{RPV-rev} \\
\hline
$|{\rm Im} (\lambda^{\prime}_{i33} \lambda^{\prime\ast}_{i11})|\cdot 
{\cal F}\left(\frac{m_b^2}{m^2_{\tilde\nu(i)}}\right)$ & 
$\leq 3.6\times 10^{-6} $ & $\leq 1.8 \times 10^{-6}$ & 
$\leq (0.8 \div 2.5) \times 10^{-8}$ & 
$|\lambda^{\prime}_{133} \lambda^{\prime}_{111}|\leq 4.5\times 10^{-5}$ 
\\
&&&& 
$|\lambda^{\prime}_{233} \lambda^{\prime}_{211}|\leq 5.4\times 10^{-3}$
\\[2mm] 
&&&& $|\lambda^{\prime}_{333} \lambda^{\prime}_{311}|\leq 1.3\times 10^{-3}$\\
\hline
$|{\rm Im} (\lambda^{\prime}_{i22} \lambda^{\prime\ast}_{i11})|\cdot 
\left(\frac{300 \, {\rm GeV}}{m_{\tilde\nu(i)}}\right)^2 $ & 
- &$\leq (0.4 \div 2.3) \times 10^{-5}$ & $\leq (0.4 \div 7.5) 
\times 10^{-7}$& 
$|\lambda^{\prime}_{122} \lambda^{\prime}_{111}|\leq 4.5\times 10^{-5}$ 
\\
&&&& $|\lambda^{\prime}_{222} \lambda^{\prime}_{211}|\leq 1.3\times 10^{-3}$
\\[2mm]
&&&& $|\lambda^{\prime}_{322} \lambda^{\prime}_{311}|\leq 1.3\times 10^{-3}$\\
\hline
$|{\rm Im} (\lambda^{\prime}_{i33} \lambda^{\prime\ast}_{i22})|\cdot 
{\cal F}\left(\frac{m_b^2}{m^2_{\tilde\nu(i)}}\right)$ & 
$\leq (1.3 \div 2.5) \times 10^{-5}$ & 
$\leq (1.3 \div 7.5) \times 10^{-5}$ & 
$\leq 2.4 \times 10^{-7}$& 
$|\lambda^{\prime}_{133} \lambda^{\prime}_{122}|\leq 4.0 \times 10^{-5}$ 
\\
&&&& $|\lambda^{\prime}_{233} \lambda^{\prime}_{222}|\leq 2.5\times 10^{-3}$
\\[2mm]
&&&& $|\lambda^{\prime}_{333} \lambda^{\prime}_{322}|\leq 3.0\times 10^{-3}$\\
\hline
\end{tabular}
\end{center} 
\end{table}

\vspace*{.5cm}

\newpage
\begin{figure}
\begin{center}
\epsfig{file=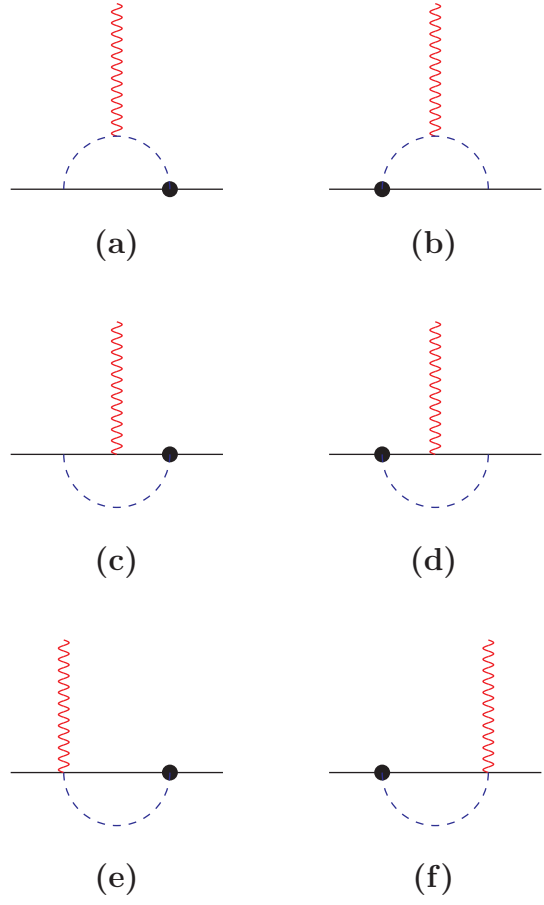, scale=1}
\end{center}

\vspace*{1cm} 
\caption{Meson-loop diagrams contributing 
to the EDMs of baryons. Solid, dashed and wiggly lines 
refer to baryons, pseudoscalar mesons and electromagnetic 
field, respectively. A CP-violating vertex is denoted by 
a black filled circle.}  
\end{figure}

\end{document}